\documentclass[prd, amsfonts, twocolumn, nofootinbib, showpacs]{revtex4}
\usepackage{graphicx, epsfig}
\usepackage{color}
\usepackage{amsmath}
\usepackage{amssymb}
\newcommand{\be}{\begin{equation}}
\newcommand{\ee}{\end{equation}}
\newcommand{\bea}{\begin{eqnarray}}
\newcommand{\eea}{\end{eqnarray}}

\newcommand{\gapp}{\mathrel{\raise.3ex\hbox{$>$}\mkern-14mu
\lower0.6ex\hbox{$\sim$}}}
\newcommand{\lapp}{\mathrel{\raise.3ex\hbox{$<$}\mkern-14mu
\lower0.6ex\hbox{$\sim$}}}
\def\bbox{{\,\lower0.9pt\vbox{\hrule \hbox{\vrule height 0.2 cm
\hskip 0.2 cm \vrule  height 0.2 cm}\hrule}\,}}

\begin{document}
\title{Collapsing objects with the same gravitational trajectory can radiate away different amount of energy}
\author{De-Chang Dai$^1$, Dejan Stojkovic$^2$}
\affiliation{$^1$ Institute of Natural Sciences, Shanghai Key Lab for Particle Physics and Cosmology, \\
and Center for Astrophysics and Astronomy, Department of Physics and Astronomy,\\
Shanghai Jiao Tong University, Shanghai 200240, China}
\affiliation{$^2$
HEPCOS, Department of Physics, SUNY, University at Buffalo, Buffalo, NY 14260-1500}

 %%%%%%%%%%%%%%%%%%%%%%%%%%%%%%%%%%%%%%%%%%%%%%%%%%%%%%%

\begin{abstract}
\widetext
We study  radiation emitted during the gravitational collapse from two different types of shells. We assume that one shell is made of dark matter and is completely transparent to the test scalar (for simplicity) field which belongs to the standard model, while the other shell is made of the standard model particles and is totally reflecting to the scalar field. These two shells have exactly the same mass, charge and angular momentum (though we set the charge and angular
momentum to zero), and therefore follow the same geodesic trajectory. However, we demonstrate that they radiate away different amount of energy during the collapse. This difference can in principle be used by an asymptotic observer to reconstruct the physical properties of the initial collapsing object other than mass, charge and angular momentum. This result has implications for the information paradox and expands the list of the type of information which can be released from a collapsing object.
\end{abstract}

%%%%%%%%%%%%%%%%%%%%%%%%%%%%%%%%%%%%%%%%%%%%%%%%%%

\maketitle

\section{Introduction}

In Einstein-Maxwell theory a stationary black hole solution is generally characterized by its mass, electric charge and angular momentum. In more general theories, some scalar field hairs have also been found \cite{Herdeiro:2014goa,Herdeiro:2015waa,Volkov:2016ehx}, and they can be considered as generalized (or Noether) charges. All additional information about the initial state of matter that formed the black hole is lost during the collapse. This includes the global charges (e.g. lepton number, baryon number, flavor \cite{Stojkovic:2005zq}), angular momentum, charge and energy distributions (as opposed to their total values which are conserved) etc.  To recover this information after the black hole is formed seems to be impossible without invoking some exotic physics. Instead of looking at the $t \rightarrow \infty$, i.e. an exact Schwarzschild solution in an asymptotically flat space-time, we can take a look at the near horizon region. Information about the initial state might be released during the collapse, since once the collapse is over there is no much one can do.
It is well known that during the collapse an object radiates away its higher multipoles and other irregularities in the so-called balding phase before a perfect spherically symmetric horizon is formed. The problem is that these are all gravitational degrees of freedom, and cannot account for other non-gravitational information content.
In \cite{Vachaspati:2006ki,Vachaspati:2007hr}, it was shown that gravitational collapse is followed by the so-called pre-Hawking radiation from the very beginning of the collapse, simply because the metric is time dependent. This radiation becomes completely thermal Hawking radiation only in $t \rightarrow \infty$ limit when the event horizon is formed. Since the collapsing object has only finite amount of mass, an asymptotic observer would never witness the formation of the horizon at $t \rightarrow \infty$. For him, the collapsing shell will slowly get converted into not-quite-thermal radiation before it reaches its own Schwarzschild radius. It was demonstrated in \cite{Saini:2015dea} that the evolution is completely unitary in such a setup.

In this paper, we also concentrate on the pre-Hawking radiation, but we are using the standard analysis as defined in \cite{Davies,1984qfcs.book.....B}. We explicitly construct an example in which two shells have exactly the same mass, charge and angular momentum (though we set the charge and angular momentum to zero for simplicity). By construct, they follow the same gravitational trajectory, however they emit different radiation during the collapse. We achieve this by giving different physical properties to the collapsing shells, other than mass, charge and angular momentum. In particular, one of the shells is completely transparent to radiation, and the other is totally reflecting. This is for example the situation where one of the shells is made of dark matter and the other of the standard model particles. If one studies emission of the standard model particles from these shells, then the dark matter shell will be completely transparent to radiation, and the standard model shell will be totally or partially reflecting.  Of course, there is a whole continuum of cases between the totally reflecting and totally transparent shells, but for the purpose of illustration, these two extremes will suffice. For simplicity, we use a spherically symmetric falling shell. In this case only s-wave scalar field is relevant, and therefore the radiation field is chosen to be a scalar field. In the realistic standard model, one could use any other field. We show that the flux of energy and power spectra of radiation emitted from these two shells is notably different, though in the limit of $t \rightarrow \infty$ the fluxes become identical. Thus, an observer studying the flux of the standard model particles from a collapsing shell could in principle tell if the shell is made of the dark or ordinary matter.

\section{The trajectory of the collapsing shell}

For our purpose, we consider a freely falling massive spherical shell. The time dependent radius of the shell is $R(\tau)$, where $\tau$ is the proper time of the observer located on the shell. The geometry outside the shell is Schwarzschild
\begin{eqnarray}
&&ds^2=-\left(1-\frac{2M}{r}\right)dt^2+\left(1-\frac{2M}{r}\right)^{-1} dr^2+r^2d\Omega\\
&&d\Omega=d\theta^2+\sin^2\theta d\phi^2  .
\end{eqnarray}
The geometry inside the shell is by the Birkhoff theorem flat Minkowski space
\begin{eqnarray}
ds^2=-dT^2+dr^2+r^2d\Omega
\end{eqnarray}
The motion of the shell can be found by matching the geometry inside and outside the shell \cite{Lightman}.
The equation of motion is given in terms of the conserved quantity $\mu$, which is just the rest mass of the shell.
\begin{equation} \label{mu}
\mu=-R\left[ (1-\frac{2M}{R}+\dot{R}^2)^{\frac{1}{2}}-(1+\dot{R}^2)^{\frac{1}{2}}\right] .
\end{equation}
Here,  $\dot{R}=\frac{dR}{d\tau}$. From Eq.~(\ref{mu}) we have
\begin{equation} \label{trajectory}
\dot{R}=\Big( \frac{M^2}{\mu^2}-1+\frac{M}{R}+\frac{\mu^2}{4R^2}\Big)^{\frac{1}{2}}
\end{equation}
Then, the proper time on the shell is given by
\begin{equation}
\tau = \int d\tau =\int \frac{dR}{\dot{R}}
\end{equation}
The time coordinate  of an asymptotic observer on the shell is
\begin{equation}
\label{out}
t = \int dt =\int \frac{\Big(1+\frac{\dot{R}^2}{1-\frac{2M}{R}}\Big)^\frac{1}{2}}{\Big(1-\frac{2M}{R}\Big)^\frac{1}{2}}d\tau
\end{equation}
The time coordinate of an observer on the shell shell is
\begin{equation}
\label{in}
T = \int dT =\int \Big(1+\dot{R}^2\Big)^\frac{1}{2}d\tau
\end{equation}

\section{Reflecting and transparent shells}

We are set to study whether two massive shells with the same gravitational trajectory can have different pre-Hawking radiation. To achieve this we consider two shells of equal mass, but one is completely transparent to a scalar field that propagates in this background, while the other one reflects the scalar field totally. The evolution of the scalar field in a curved background outside the shell is described by
\begin{equation}
\Box \phi =0
\end{equation}
where the $\Box$ operator is covariant. Inside the shell, the $\Box$ operator is Minkowski.  Because of the spherical symmetry, as usual, we simplify the discussion and focus on a $1+1$ dimensional scalar field, $\phi (t,r)$, which satisfies the wave equation

\begin{eqnarray}
&& \partial_t^2 \phi- \partial_{r^*}^2\phi =0 \mbox{, for $r>R$}\\
&& \partial_T^2 \phi- \partial_r^2 \phi =0 \mbox{, for $r<R$}
\end{eqnarray}
Here $r^*=\int \frac{dr}{1-\frac{2M}{r}}$ is the usual tortoise coordinate. The trajectory of the spherical shell is given by Eq.~(\ref{trajectory}), and it is the same for both shells since they have the same mass (and carry no charge nor angular momentum). There are two types of solutions to the wave equation for $r> R$, i.e. $f(t\pm r^*)$. The function $f(t-r^*)$ represents a wave moving to the right, while $f(t+r^*)$ represents a wave moving to the left. When a plane wave is propagating inward toward the origin, it is considered as an ingoing mode and can be written as
\begin{equation}
\phi_{\rm in} \sim \exp(-i\omega v)
\end{equation}
where we defined the ingoing and outgoing null coordinates $v=t+r^*$ and $u=t-r^*$.
When the ingoing mode passes through the center, it starts propagating outward (away from the center), and it becomes an outgoing mode. The form of wave function is the same as before, but its argument must be a function of an outgoing coordinate $f(u)$, i.e.
\begin{equation}
\phi_{\rm out} \sim \exp(-i\omega p(u)) ,
\end{equation}
where $p(u)$ is a function of the coordinate $u$.

The shells in our discussion here are massive, which is different from the massless shells discussed in usual cases. The massless scalar field is moving faster than the shell and will pass through or be reflected by the matter on the shell.

\begin{figure}
   \centering
\includegraphics[width=5cm]{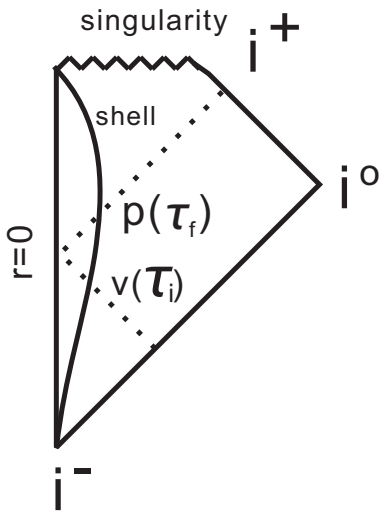}
\caption{Penrose diagram for the transparent collapsing shell. The mode crosses the shell at some initial time $\tau_i$, passes through the center, and crosses again at some final time $\tau_f$.}
\label{Tran}
\end{figure}

We now consider the transparent shell first. While the shell is collapsing, the incoming scalar field mode passes through the shell and reaches the center of the shell. Once it passes through the center, it becomes an outgoing mode. As shown in Fig.~\ref{Tran}, the mode crosses the shell at some initial time $\tau_i$, passes through the center, and crosses again at some final time $\tau_f$. Since the field moves at the speed of light, $\tau_i$ and $\tau_f$ must satisfy the condition
\begin{equation}
\label{con1}
R(\tau_f)+R(\tau_i)=T(\tau_f)-T(\tau_i) .
\end{equation}
A scalar field coming from $R(\tau_i)$ passing through the center and arriving to  $R(\tau_f)$ travels the distance
$R(\tau_i)+R(\tau_f)$. An inside observer measures the time of this process $T(\tau_f)-T(\tau_i)$. Since the massless field travels at the speed of light, we arrive to Eq.~(\ref{con1}).

\begin{figure}
   \centering
\includegraphics[width=9cm]{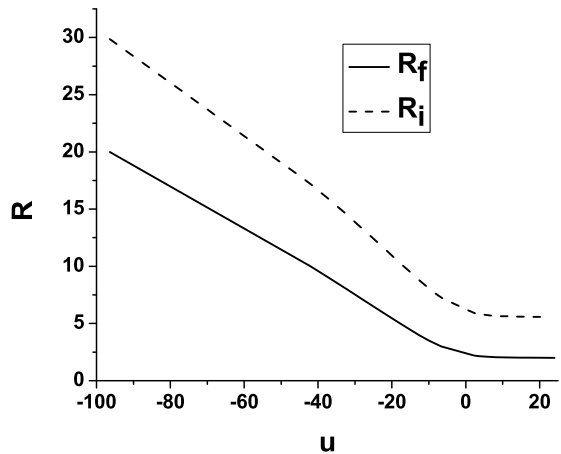}
\caption{In this figure we set $u=t(\tau_f)-r^*(\tau_f )$ and $M =1$. $R_i$ is the radius of the shell when the scalar field mode crosses it for the first time at $\tau_i$ on the way inside the shell, and $R_f$ is the radius when the mode crosses it for the second time on the way out $\tau_f$. }
\label{radius}
\end{figure}

The shell's radius at the moments of these two crossings is shown in Fig.~\ref{radius}.

When the wave mode comes out of the shell we have $u=t(\tau_f)-r^*(\tau_f)$, but $p=v(\tau_i)=t(\tau_i)+r^*(\tau_i)$. The function $p$ can be written in terms of the variable $u$ with the help of Eq.~(\ref{con1}).

\begin{figure}
   \centering
\includegraphics[width=5cm]{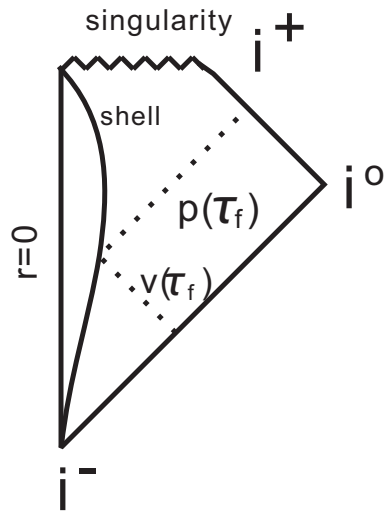}
\caption{Penrose diagram for the totally reflecting collapsing  shell. The scalar field ingoing mode is reflected by the shell at time $\tau_f$.}
\label{Ref}
\end{figure}

 The totally reflecting case is shown in Fig.~\ref{Ref}. An ingoing mode is reflected by the shell immediately at $\tau_f$. Therefore in this case $p=v(\tau_f)=t(\tau_f)+r^* (\tau_f)$ and $u=t(\tau_f)-r^* (\tau_f)$. The function $p$ can now be written in terms of the variable $u$  by replacing $\tau_f$ with $u$.

We will now study the energy flux coming from the shell in these two cases.

\section{Energy flux and power spectrum}

The renormalized stressed-energy tensor for a massless scalar field was computed in terms of $p(u)$ by Fulling and Davies\cite{Davies,1984qfcs.book.....B}. In a $1+1$ dimensional spacetime, it can be written as
\begin{eqnarray}
T_{uu}&=&\frac{1}{24\pi}\Bigg( -\frac{M}{r^3}+\frac{3}{2}\frac{M^2}{r^2}+\frac{3}{2}\Big(\frac{p''}{p'}\Big)^2-\frac{p'''}{p'}\Bigg)\\
T_{uv}&=&-\frac{1}{24\pi}\Big( 1-\frac{2M}{r}\Big) \frac{M}{r^3}\\
T_{vv}&=&\frac{1}{24\pi}\Big( -\frac{M}{r^3}+\frac{3}{2}\frac{M^2}{r^2}\Big)
\end{eqnarray}
Primes denote derivatives with respect to the coordinate $u$. As $r\rightarrow \infty$, only $T_{uu}$ survives, which is the radiated energy flux we are looking for. To get concrete numerical results, for simplicity we set $\mu =M =1$. We can find the coordinates $t$ and $T$ from Eqs.(\ref{out}) and (\ref{in}). There will be an arbitrary integration constant in these two integrals, but their values will not affect the result, so we set them to zero.

\begin{figure}
   \centering
\includegraphics[width=9cm]{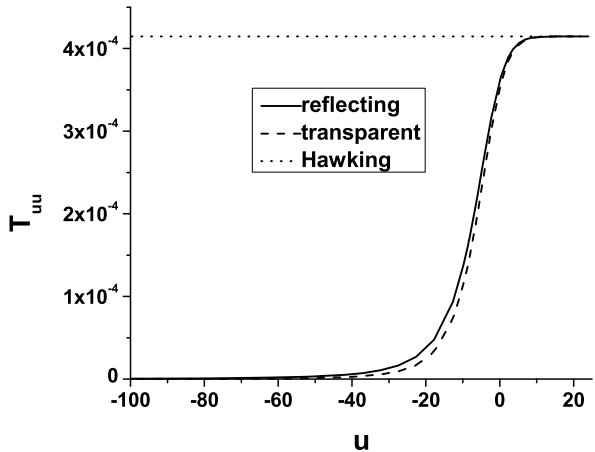}
\caption{The black curve represents the energy flux $T_{uu}$ at $r\rightarrow \infty$ for the totally reflecting shell. The dashed curve represents the energy flux $T_{uu}$ at $r\rightarrow \infty$ for the totally transparent shell. The doted curve is the Hawking radiation for a static black hole, which is just $\frac{1}{768\pi}$. At late times, the two shells match the Hawking radiation from a pre-existing black hole, but the fluxes are substantially different during the collapse. }
\label{flux}
\end{figure}

\begin{figure}
   \centering
\includegraphics[width=9cm]{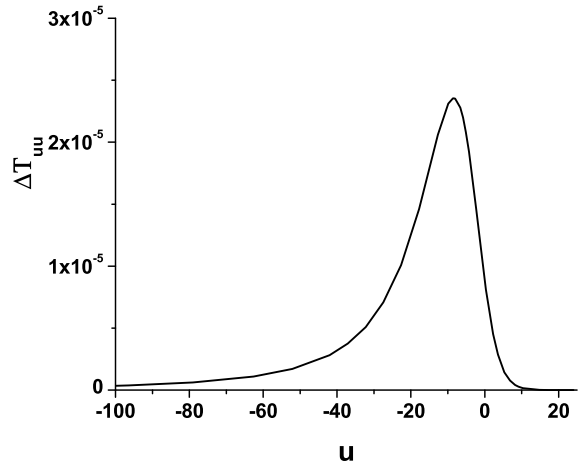}
\caption{The black curve represents the energy flux difference between totally reflecting shell and transparent shell, $\Delta T_{uu}$, at $r\rightarrow \infty$. At first,  at $u\rightarrow -\infty$, there is no radiation and therefore no difference, so $\Delta T_{uu}\rightarrow 0$. $\Delta T_{uu}$ achieves its maximal at $u\approx -8.3$. After that, radiation from both shells becomes very close to Hawking radiation, and $\Delta T_{uu}$ goes back to zero.  }
\label{diff}
\end{figure}

We plot the term $T_{uu}$ as a function of $r$ in Fig.~\ref{flux}. It is obvious that these two shells emits different fluxes as seen at infinity. The reflecting shell has stronger pre-Hawking radiation and emits more energy than the transparent shell. The difference is maximal around the value of the
time parameter $u\approx -8.3$ (see Fig.~\ref{diff}).

Obviously, the totally reflecting shell emits more energy than a transparent shell (Fig.~\ref{flux}). This difference is coming from the interaction between the shell and the $\phi$ field in vacuum. Since the reflecting shell can interact with $\phi$ field, it affects the vacuum stronger than the transparent shell does.  This interaction that does not exist in the transparent case may be interpreted as that the fact that reflecting shell contains more information than the transparent one. This information then needs to be released before a static featureless black hole is formed. At late time, in the $t\rightarrow \infty$ limit,  radiation from both shells is indistinguishable and matches the radiation from a static black hole. In this limit, all the black holes with the same mass emit the same radiation, no matter what they are made of (e.g. dark vs. ordinary matter). As the main result of our analysis, we state that pre-Hawking radiation from two shells is different, even though the gravitational trajectories are the same.

\section{Conclusion}

In Einstein-Maxwell theory, physical properties of a black hole are completely determined by its mass, electric charge and angular momentum. There is no additional information left after a black hole is formed. Therefore  black holes made of matter with different properties (other than mass, charge and angular momentum) emit exactly the same Hawking radiation. Then one cannot recover any additional information with conventional physics if only Hawking radiation from a static black hole is considered.

It is well known that a collapsing object can shed its higher multipole moments in the so-called balding phase before reaching a perfect spherically symmetric form. However it is important to extend this balding phase to other non-gravitational degrees of freedom.

In this paper, we considered a specific problem whether a collapsing object made of dark matter radiates away energy in the same way as an object made of the standard model particles. If an asymptotic observer registers the flux of the standard model particles (for a simplicity a standard model scalar field), then the collapsing shell made of dark matter will be almost completely transparent to the scalar field due to the very weak interactions between the dark matter and the standard model. In contrast, the shell made of the standard model particles will be almost completely reflecting.
These two shells have the same mass, charge and angular momentum, therefore, their trajectories dictated by gravity will be exactly the same. Both of these shells
will give rise to pre-Hawking radiation, so the question is whether this radiation will be identical for both shells
or not. We calculate the trajectory for the massive shell, and then the components of energy momentum tensor in
a $(1+1)$-dim spherically symmetric space-time defined by $(t,r)$. The scalar field ingoing mode passes through
the transparent shell, reaches the center and then reappears on the other side of the shell as an outgoing mode.
In contrast, scalar field ingoing mode gets reflected back from the reflecting shell and becomes an outgoing mode
immediately. This difference is sufficient to give different amount and power spectrum of radiation. The main results are shown in Fig.~\ref{flux}. The shells indeed radiate in a different way before becoming a black hole.
The reflecting shell emits more energy than the transparent shell. This difference is caused by the interaction
between the shell and the vacuum. At late time, in the $t\rightarrow \infty$ limit, radiation from both shells is indistinguishable and matches the radiation from a static black hole.

We therefore demonstrated that pre-Hawking radiation can be used in principle to recover some physical quantities of the collapsing matter. In this concrete example, an observer can tell if an object was  made of ordinary or dark matter.
This information is released out during the collapse and it should be part of the balding phase in the black hole formation. One should not confuse this result with the usual balding
phase, which only includes quantities that affect gravity, i.e. mass, charge, angular momentum and position.
This is an extra balding effect which reveals other physical quantities of the collapsing object.

\begin{acknowledgments}
D.C Dai was supported by the National Science Foundation of China (Grant No. 11433001 and 11447601), National Basic Research Program of China (973 Program 2015CB857001), No.14ZR1423200 from the Office of Science and Technology in Shanghai Municipal Government, the key laboratory grant from the Office of Science and Technology in Shanghai Municipal Government (No. 11DZ2260700) and  the Program of Shanghai Academic/Technology Research Leader under Grant No. 16XD1401600. DS partially supported by the US National Science Foundation, under Grant No. PHY-1417317.

\end{acknowledgments}

\end{document}